\documentclass[12pt,a4]{article}

\usepackage{natbib}
\usepackage{color,graphicx}
\usepackage{makeidx}
\usepackage{csquotes}
\usepackage{array}
\usepackage{graphicx}
 \usepackage[onehalfspacing]{setspace}
\usepackage{amsmath, amsthm, amsfonts, bm}
\usepackage{amssymb}
\usepackage{rotating}
\usepackage{color}
\usepackage[]{algorithm2e}
\usepackage{pbox}
\usepackage{longtable}
\usepackage{subcaption}
\usepackage{hyperref} 
\usepackage{listings}
\usepackage{lscape}
\usepackage{rotate}
\usepackage{booktabs}
\usepackage{authblk}
\usepackage[labelfont=bf]{caption}

\usepackage[usenames,dvipsnames]{xcolor}

\title{Minnesota BART\thanks{\scriptsize
\textit{Corresponding author}: Pedro Lima. The University of Texas at Austin. \textit{Email}: \href{mailto:plima@utexas.edu}{plima@utexas.edu}. We would like to thank the participants of the 2024 European Seminar on Bayesian Econometrics (ESOBE), VII COBAL, and XVII EBEB for their many constructive comments and useful suggestions. Hedibert Lopes also acknowledges partial
financial support from FAPESP grants 2023/02538-0 and 2024/01027-4.}}

\author[1]{Pedro A. Lima}
\author[1]{Carlos M. Carvalho}
\author[2]{Hedibert F. Lopes}
\author[1]{Andrew Herren}
\affil[1]{The University of Texas at Austin}
\affil[2]{Insper}

\date{\today}

\begin{document}

\onehalfspacing



\newcommand{\widgraph}[2]{\includegraphics[keepaspectratio,width=#1]{#2}}

\renewcommand{\arraystretch}{1.4}


\newtheorem*{remark}{Remark}

\newtheorem{assumption}{Assumption}

\newtheorem{lemma}{Lemma}
\newtheorem{example}{Example}
\newtheorem{theorem}{Theorem}
\newtheorem{proposition}{Proposition}
\newtheorem{definition}{Definition}
\newtheorem{corollary}{Corollary}
\newenvironment{assumptionprime}[1]
  {\renewcommand{\theassumption}{\ref{#1}$'$}%
   \addtocounter{assumption}{-1}%
   \begin{assumption}}
  {\end{assumption}}

\newcommand{\Real}{\mathbb{R}}
\newcommand{\bbP}{\mathbb{P}}
\newcommand{\bbE}{\mathbb{E}}
\newcommand{\var}{\mathrm{Var}}

\newcommand{\softmax}{\mathrm{Softmax}}
\newcommand{\sigmoid}{\mathrm{sigmoid}}
\newcommand{\gelu}{\mathrm{GELU}}
\newcommand{\relu}{\mathrm{ReLU}}

\newcommand{\dint}{\mathrm{d}}

\newcommand{\red}[1]{\textcolor{red}{#1}}
\newcommand{\blue}[1]{\textcolor{blue}{#1}}
\newcommand{\black}[1]{\textcolor{black}{#1}}
\newcommand{\cyan}[1]{\textcolor{cyan}{#1}}
\newcommand{\magenta}[1]{\textcolor{magenta}{#1}}
\newcommand{\pink}[1]{\textcolor{pink}{#1}}

\def\st{{\em s.t.~}}
\def\ie{{\em i.e.,~}}
\def\eg{{\em e.g.,~}}
\def\cf{{\em cf.,~}}
\def\ea{{\em et al.~}}
\newcommand{\iid}{i.i.d.}
\newcommand{\wrt}{w.r.t.}

\newcommand{\dboijn}{\Delta \beta_{1ij}^{n}}
\newcommand{\dbzijn}{\Delta \beta_{0ij}^{n}}
\newcommand{\dtn}{\Delta \tau^n}
\newcommand{\daijn}{\Delta a_{ij}^{n}}
\newcommand{\dbijn}{\Delta b_{ij}^{n}}
\newcommand{\deijn}{\Delta \eta_{ij}^{n}}

\newcommand{\dboin}{\Delta \beta_{1i}^{n}}
\newcommand{\dbboin}{\Delta \bar{\beta}_{1i}^{n}}
\newcommand{\dbzin}{\Delta \beta_{0i}^{n}}
\newcommand{\dbbzin}{\Delta \bar{\beta}_{0i}^{n}}

\newcommand{\dain}{\Delta a_{i}^{n}}
\newcommand{\dbin}{\Delta b_{i}^{n}}
\newcommand{\dein}{\Delta \eta_{i}^{n}}
\newcommand{\dbein}{\Delta \bar{\eta}_{i}^{n}}

\newcommand{\norm}[1]{\|#1\|}

\newcommand{\boin}{\beta_{1i}^n}
\newcommand{\bzin}{\beta_{0i}^n}
\newcommand{\ain}{a_i^n}
\newcommand{\bin}{b_i^n}
\newcommand{\ein}{\eta_i^n}

\newcommand{\boonen}{\beta_{11}^n}
\newcommand{\bzonen}{\beta_{01}^n}
\newcommand{\aonen}{a_1^n}
\newcommand{\bonen}{b_1^n}
\newcommand{\eonen}{\eta_1^n}

\newcommand{\boj}{\beta_{1j}^*}
\newcommand{\bzj}{\beta_{0j}^*}
\newcommand{\aj}{a_j^*}
\newcommand{\bj}{b_j^*}
\newcommand{\ej}{\eta_j^*}

\newcommand{\bboi}{\bar{\beta}_{1i}}
\newcommand{\bbzi}{\bar{\beta}_{0i}}
\newcommand{\bei}{\bar{\eta_i}}

\newcommand{\boi}{\beta_{1i}^*}
\newcommand{\bzi}{\beta_{0i}^*}
\newcommand{\ai}{a_i^*}
\newcommand{\bi}{b_i^*}
\newcommand{\ei}{\eta_i^*}

\newcommand{\boone}{\beta_{11}^*}
\newcommand{\bzone}{\beta_{01}^*}
\newcommand{\aone}{a_1^*}
\newcommand{\bone}{b_1^*}
\newcommand{\eone}{\eta_1^*}

\newcommand{\cjp}{c_{j'}^0}
\newcommand{\gjp}{\Gamma_{j'}^0}
\newcommand{\ajp}{a_{j'}^0}
\newcommand{\bjp}{b_{j'}^0}
\newcommand{\ejp}{\eta_{j'}^0}

\newcommand{\zerod}{{0}_d}
\newcommand{\zeroq}{\mathbf{0}_q}

\newcommand{\kbar}{\bar{k}}
\newcommand{\ktilde}{\tilde{k}}
\newcommand{\Dtilde}{\widetilde{D}}

\newcommand{\brj}{\bar{r}_j}
\newcommand{\trj}{\tilde{r}(|\mathcal{A}_j|)}
\newcommand{\trjp}{\tilde{r}(|\mathcal{A}_{j'}|)}

\newcommand{\brone}{\bar{r}_1}
\newcommand{\pizeroone}{\pi_{1}^{0}}
\newcommand{\dboione}{\Delta \beta_{1i1}^{n}}
\newcommand{\dtone}{\Delta \tau^{n}}

\newcommand{\daione}{\Delta a_{i1}^{n}}
\newcommand{\dbione}{\Delta b_{i1}^{n}}
\newcommand{\deione}{\Delta \eta_{i1}^{n}}

\newcommand{\trone}{\tilde{r}(|\mathcal{A}_1|)}
\newcommand{\trs}{\tilde{r}(|\mathcal{A}_{j^*}|)}

\newcommand{\normf}[1]{\|#1\|_{L_2(\mu)}}

\def\reta{{\textnormal{$\eta$}}}
\def\ra{{\textnormal{a}}}
\def\rb{{\textnormal{b}}}
\def\rc{{\textnormal{c}}}
\def\rd{{\textnormal{d}}}
\def\re{{\textnormal{e}}}
\def\rf{{\textnormal{f}}}
\def\rg{{\textnormal{g}}}
\def\rh{{\textnormal{h}}}
\def\ri{{\textnormal{i}}}
\def\rj{{\textnormal{j}}}
\def\rk{{\textnormal{k}}}
\def\rl{{\textnormal{l}}}

\def\rn{{\textnormal{n}}}
\def\ro{{\textnormal{o}}}
\def\rp{{\textnormal{p}}}
\def\rq{{\textnormal{q}}}
\def\rr{{\textnormal{r}}}
\def\rs{{\textnormal{s}}}
\def\rt{{\textnormal{t}}}
\def\ru{{\textnormal{u}}}
\def\rv{{\textnormal{v}}}
\def\rw{{\textnormal{w}}}
\def\rx{{\textnormal{x}}}
\def\ry{{\textnormal{y}}}
\def\rz{{\textnormal{z}}}

\def\rvepsilon{{\mathbf{\epsilon}}}
\def\rvtheta{{\mathbf{\theta}}}
\def\rva{{\mathbf{a}}}
\def\rvb{{\mathbf{b}}}
\def\rvc{{\mathbf{c}}}
\def\rvd{{\mathbf{d}}}
\def\rve{{\mathbf{e}}}
\def\rvf{{\mathbf{f}}}
\def\rvg{{\mathbf{g}}}
\def\rvh{{\mathbf{h}}}
\def\rvu{{\mathbf{i}}}
\def\rvj{{\mathbf{j}}}
\def\rvk{{\mathbf{k}}}
\def\rvl{{\mathbf{l}}}
\def\rvm{{\mathbf{m}}}
\def\rvn{{\mathbf{n}}}
\def\rvo{{\mathbf{o}}}
\def\rvp{{\mathbf{p}}}
\def\rvq{{\mathbf{q}}}
\def\rvr{{\mathbf{r}}}
\def\rvs{{\mathbf{s}}}
\def\rvt{{\mathbf{t}}}
\def\rvu{{\mathbf{u}}}
\def\rvv{{\mathbf{v}}}
\def\rvw{{\mathbf{w}}}
\def\rvx{{\mathbf{x}}}
\def\rvy{{\mathbf{y}}}
\def\rvz{{\mathbf{z}}}

\def\erva{{\textnormal{a}}}
\def\ervb{{\textnormal{b}}}
\def\ervc{{\textnormal{c}}}
\def\ervd{{\textnormal{d}}}
\def\erve{{\textnormal{e}}}
\def\ervf{{\textnormal{f}}}
\def\ervg{{\textnormal{g}}}
\def\ervh{{\textnormal{h}}}
\def\ervi{{\textnormal{i}}}
\def\ervj{{\textnormal{j}}}
\def\ervk{{\textnormal{k}}}
\def\ervl{{\textnormal{l}}}
\def\ervm{{\textnormal{m}}}
\def\ervn{{\textnormal{n}}}
\def\ervo{{\textnormal{o}}}
\def\ervp{{\textnormal{p}}}
\def\ervq{{\textnormal{q}}}
\def\ervr{{\textnormal{r}}}
\def\ervs{{\textnormal{s}}}
\def\ervt{{\textnormal{t}}}
\def\ervu{{\textnormal{u}}}
\def\ervv{{\textnormal{v}}}
\def\ervw{{\textnormal{w}}}
\def\ervx{{\textnormal{x}}}
\def\ervy{{\textnormal{y}}}
\def\ervz{{\textnormal{z}}}

\def\rmA{{\mathbf{A}}}
\def\rmB{{\mathbf{B}}}
\def\rmC{{\mathbf{C}}}
\def\rmD{{\mathbf{D}}}
\def\rmE{{\mathbf{E}}}
\def\rmF{{\mathbf{F}}}
\def\rmG{{\mathbf{G}}}
\def\rmH{{\mathbf{H}}}
\def\rmI{{\mathbf{I}}}
\def\rmJ{{\mathbf{J}}}
\def\rmK{{\mathbf{K}}}
\def\rmL{{\mathbf{L}}}
\def\rmM{{\mathbf{M}}}
\def\rmN{{\mathbf{N}}}
\def\rmO{{\mathbf{O}}}
\def\rmP{{\mathbf{P}}}
\def\rmQ{{\mathbf{Q}}}
\def\rmR{{\mathbf{R}}}
\def\rmS{{\mathbf{S}}}
\def\rmT{{\mathbf{T}}}
\def\rmU{{\mathbf{U}}}
\def\rmV{{\mathbf{V}}}
\def\rmW{{\mathbf{W}}}
\def\rmX{{\mathbf{X}}}
\def\rmY{{\mathbf{Y}}}
\def\rmZ{{\mathbf{Z}}}

\def\ermA{{\textnormal{A}}}
\def\ermB{{\textnormal{B}}}
\def\ermC{{\textnormal{C}}}
\def\ermD{{\textnormal{D}}}
\def\ermE{{\textnormal{E}}}
\def\ermF{{\textnormal{F}}}
\def\ermG{{\textnormal{G}}}
\def\ermH{{\textnormal{H}}}
\def\ermI{{\textnormal{I}}}
\def\ermJ{{\textnormal{J}}}
\def\ermK{{\textnormal{K}}}
\def\ermL{{\textnormal{L}}}
\def\ermM{{\textnormal{M}}}
\def\ermN{{\textnormal{N}}}
\def\ermO{{\textnormal{O}}}
\def\ermP{{\textnormal{P}}}
\def\ermQ{{\textnormal{Q}}}
\def\ermR{{\textnormal{R}}}
\def\ermS{{\textnormal{S}}}
\def\ermT{{\textnormal{T}}}
\def\ermU{{\textnormal{U}}}
\def\ermV{{\textnormal{V}}}
\def\ermW{{\textnormal{W}}}
\def\ermX{{\textnormal{X}}}
\def\ermY{{\textnormal{Y}}}
\def\ermZ{{\textnormal{Z}}}

\def\vzero{{\bm{0}}}
\def\vone{{\bm{1}}}
\def\vmu{{\bm{\mu}}}
\def\vdelta{{\bm{\delta}}}
\def\veta{{\bm{\eta}}}
\def\vtheta{{\bm{\theta}}}
\def\vlambda{{\bm{\lambda}}}
\def\va{{\bm{a}}}
\def\vb{{\bm{b}}}
\def\vc{{\bm{c}}}
\def\vd{{\bm{d}}}
\def\ve{{\bm{e}}}
\def\vf{{\bm{f}}}
\def\vg{{\bm{g}}}
\def\vh{{\bm{h}}}
\def\vi{{\bm{i}}}
\def\vj{{\bm{j}}}
\def\vk{{\bm{k}}}
\def\vl{{\bm{l}}}
\def\vm{{\bm{m}}}
\def\vn{{\bm{n}}}
\def\vo{{\bm{o}}}
\def\vp{{\bm{p}}}
\def\vq{{\bm{q}}}
\def\vr{{\bm{r}}}
\def\vs{{\bm{s}}}
\def\vt{{\bm{t}}}
\def\vu{{\bm{u}}}
\def\vv{{\bm{v}}}
\def\vw{{\bm{w}}}
\def\vx{{\bm{x}}}
\def\vy{{\bm{y}}}
\def\vz{{\bm{z}}}

\def\evalpha{{\alpha}}
\def\evbeta{{\beta}}
\def\evepsilon{{\epsilon}}
\def\evlambda{{\lambda}}
\def\evomega{{\omega}}
\def\evmu{{\mu}}
\def\evpsi{{\psi}}
\def\evsigma{{\sigma}}
\def\evtheta{{\theta}}
\def\eva{{a}}
\def\evb{{b}}
\def\evc{{c}}
\def\evd{{d}}
\def\eve{{e}}
\def\evf{{f}}
\def\evg{{g}}
\def\evh{{h}}
\def\evi{{i}}
\def\evj{{j}}
\def\evk{{k}}
\def\evl{{l}}
\def\evm{{m}}
\def\evn{{n}}
\def\evo{{o}}
\def\evp{{p}}
\def\evq{{q}}
\def\evr{{r}}
\def\evs{{s}}
\def\evt{{t}}
\def\evu{{u}}
\def\evv{{v}}
\def\evw{{w}}
\def\evx{{x}}
\def\evy{{y}}
\def\evz{{z}}

\def\mA{{\bm{A}}}
\def\mB{{\bm{B}}}
\def\mC{{\bm{C}}}
\def\mD{{\bm{D}}}
\def\mE{{\bm{E}}}
\def\mF{{\bm{F}}}
\def\mG{{\bm{G}}}
\def\mH{{\bm{H}}}
\def\mI{{\bm{I}}}
\def\mJ{{\bm{J}}}
\def\mK{{\bm{K}}}
\def\mL{{\bm{L}}}
\def\mM{{\bm{M}}}
\def\mN{{\bm{N}}}
\def\mO{{\bm{O}}}
\def\mP{{\bm{P}}}
\def\mQ{{\bm{Q}}}
\def\mR{{\bm{R}}}
\def\mS{{\bm{S}}}
\def\mT{{\bm{T}}}
\def\mU{{\bm{U}}}
\def\mV{{\bm{V}}}
\def\mW{{\bm{W}}}
\def\mX{{\bm{X}}}
\def\mY{{\bm{Y}}}
\def\mZ{{\bm{Z}}}
\def\mBeta{{\bm{\beta}}}
\def\mPhi{{\bm{\Phi}}}
\def\mLambda{{\bm{\Lambda}}}
\def\mSigma{{\bm{\Sigma}}}
\def\mOmega{{\bm{\Omega}}}
\newcommand{\error}{\varepsilon} 
\newcommand{\verror}{\bm{\error}}

\newcommand{\cF}{\mathcal{F}}
\newcommand{\cA}{\mathcal{A}}
\newcommand{\cB}{\mathcal{B}}
\newcommand{\cM}{\mathcal{M}}
\newcommand{\cD}{\mathcal{D}}
\newcommand{\cT}{\mathcal{T}}
\newcommand{\cP}{\mathcal{P}}
\newcommand{\cN}{\mathcal{N}}
\newcommand{\cL}{\mathcal{L}}


\newcommand{\LPS}{{\mbox{\rm LPDS}}}
\newcommand{\LPSo}[1]{\LPS^{\star}_{#1}}
\newcommand{\thc}{\vartheta}
\newcommand{\thmod}{{\mathbf{\boldsymbol{\thc}}}}

\maketitle 

\begin{abstract}

\noindent {\small Vector autoregression (VAR) models are widely used for forecasting and macroeconomic analysis, yet they remain limited by their reliance on a linear parameterization. Recent research has introduced nonparametric alternatives, such as Bayesian additive regression trees (BART), which provide flexibility without strong parametric assumptions.  However, existing BART-based frameworks do not account for time dependency or allow for sparse estimation in the construction of regression tree priors, leading to noisy and inefficient high-dimensional representations. This paper introduces a sparsity-inducing Dirichlet hyperprior on the regression tree's splitting probabilities, allowing for automatic variable selection and high-dimensional VARs. Additionally, we propose a structured shrinkage prior that decreases the probability of splitting on higher-order lags, aligning with the Minnesota prior’s principles. Empirical results demonstrate that our approach improves predictive accuracy over the baseline BART prior and Bayesian VAR (BVAR), particularly in capturing time-dependent relationships and enhancing density forecasts. These findings highlight the potential of developing domain-specific nonparametric methods in macroeconomic forecasting.}
\end{abstract}

\onehalfspacing
\section{Introduction}

This paper introduces a multivariate Bayesian additive regression tree (BART) model for macroeconomic forecasting that incorporates structured priors to allow for variable selection and account for time dependency. We extend the standard BART framework to allow for a sparse vector autoregressions (VARs) estimation by introducing a sparsity-inducing Dirichlet hyperprior on the regression tree’s splitting probabilities. This allows for automatic variable selection, reducing overfitting and improving computational efficiency in large-scale models.  

Additionally, we propose a structured shrinkage prior that decreases the probability of splitting on higher-order lags, aligning with the Minnesota prior’s principles. This addresses a fundamental limitation of existing BART-based VAR models, which fail to incorporate economic constraints on lag selection and ignore temporal dependencies. We also analyze how different levels of our shrinkage parameter affects the splitting probabilities, results demonstrate that higher values of the parameter lead to a more gradual decay in posterior inclusion probabilities, preserving the influence of lags and cross-lags for a longer range. This choice also can affect the forecasting performance of the model. By integrating these two enhancements, our approach preserves the flexibility of BART while imposing meaningful structure, leading to improved interpretability and forecasting accuracy.  

We evaluate our model through an empirical application to U.S. macroeconomic forecasting. Compared to standard BART and Bayesian VAR (BVAR) models, our approach improves predictive accuracy, particularly in capturing time-dependent relationships and higher-order moments of the predictive distribution. In particular, we show that our method enhances density forecasts for key macroeconomic variables such as the Federal Funds Rate and inflation. These findings highlight the potential of structured nonparametric methods for macroeconomic forecasting.  

Vector autoregression (VAR) models have been widely used for forecasting and structural analysis of macroeconomic variables (\cite{doan1984forecasting}; \cite{litterman1986forecasting}; \cite{banbura2010large}; \cite{koop2013forecasting}; \cite{carriero2019large}; \cite{kastner2020sparse}). However, as the number of time series included in the model increases, the number of parameters grows quadratically, leading to concerns about overparameterization and in-sample overfitting. To address these challenges, the Bayesian literature on VARs has developed various shrinkage prior specifications.  

Despite these advancements, most Bayesian VAR models still assume a linear relationship between endogenous variables and their lags. While macroeconomic relationships are often stable over time, allowing a linear approximation to fit the data reasonably well, this assumption can break down during shocks that alter the economy’s dynamics (\cite{huber2023nowcasting}). Failing to account for such events can result in poor out-of-sample performance and misinterpretation of impulse response functions.  

More recently, nonparametric approaches such as Bayesian additive regression trees (BART) have gained attention as a flexible alternative (\cite{chipman2010bart}). BART uses regression trees as weak learners, allowing for complex relationships to be modeled without strong parametric assumptions. However, existing frameworks (\cite{huber2022inference}; \cite{clark2023tail}) do not accommodate high-dimensional settings or account for time dependency in the construction of regression tree priors. Our paper directly addresses these gaps.  

The remainder of the paper is structured as follows. Section 2 introduces the multivariate BART model and provides a necessary introduction to the BART framework. Section 3 develops the prior construction. Section 4 details the prior setup and posterior computation. Sections 5 and 6 present our empirical results: Section 5 discusses the dataset and provides in- and out-of-sample model evidence, while Section 6 focuses on macroeconomic forecasting results. The final section summarizes and concludes.

\section{Tree Based Vector Autoregression model}
\subsection{The Model}

 Define $\vy_t = \left(y_{1t}, \dots, y_{nt}\right)'$ as a vector of endogenous variables of dimension $n\times 1$. Define $\vx_t = \left(\vy_{t-1}', \dots, \vy_{t-p}'\right)'$ a $k(= np)$ dimensional vector of covariates, where $p$ is the number of lags. We also define $\mG(\vx_t)=\left(g_1(\vx_t), \ldots, g_n(\vx_t)\right)'$ as a n-dimensional vector of non-parametric functions :
\begin{eqnarray}
\mathbf{y_t} &=& \mG(\vx_t) + \verror_t, \quad \verror_t \sim \mathcal{N}(0,\mSigma_t), \quad \text{for } t =1, \dots , T
\end{eqnarray}

\noindent
Each function in the vector $\mG(\cdot)$, where $\mG: \Real^{k}\rightarrow \Real^{n}$, hence $g\left(\vx_t\right): \Real^{k} \rightarrow \Real$ will be approximated by a Bayesian Additive Regression Trees (BART) model, which is discussed in detail in Section \ref{sec:bart}.

 Research on large Bayesian vector autoregressions (VARs) shows evidence that stochastic volatility specifications are well supported by the data (\cite{carriero2016common}; \cite{koop2013forecasting}; \cite{chan2020large}) and achieving precise density forecasts. Recent work by \cite{chan2023comparing} demonstrates that the factor stochastic volatility and Cholesky stochastic volatility specifications outperform the standard common stochastic volatility model.

 Therefore, following \cite{huber2022inference}, the conditional covariance structure is specified as factor stochastic volatility (FSV) (\cite{pitt1999time}; \cite{aguilar2000bayesian}; \cite{chib2006analysis}; \cite{lopes2007factor}; \cite{kastner2020sparse}). More precisely, the error term is decomposed as:

\begin{eqnarray}
\label{cov}
\verror_t  = \mLambda\vf_t + \veta_t, 
\end{eqnarray}

\noindent where $\vf_t = \left( f_{1,t}, \dots, f_{r,t} \right)$ is a $r\times1$ vector of latent factors and $\mLambda$ is the associated $n\times r$ factor loading matrix. This factor specification is not identified. The latent factors and the idiosyncratic errors are assumed to be independent and jointly Gaussian:

\begin{equation}
\begin{pmatrix}
\veta_t \\
\vf_t
\end{pmatrix}
\sim \mathcal{N} \left(
\begin{pmatrix}
\mathbf{0} \\
\mathbf{0}
\end{pmatrix},
\begin{bmatrix}
\mOmega_t & \mathbf{0} \\
\mathbf{0} & \mH_t
\end{bmatrix}
\right).
\end{equation}

\noindent where $\mOmega_t = diag\left( e^{h_{1,t}}, \dots, e^{h_{n,t}} \right)$ and $\mH_t = diag\left( e^{h_{n+1,t}}, \dots, e^{h_{n+r,t}} \right)$ are diagonal matrices. The evolution of the log-variance process for $i = 1, \dots, n+r$ is defined as:
\begin{eqnarray}
\label{sv}
    h_{i,t} = \mu_i + \phi_i(h_{i,t} - \mu_i) + \eta_{i,t}^{h}, \quad \eta_{i,t}^{h} \sim \mathcal{N}\left(0, \sigma_{i}^{2}\right)
\end{eqnarray}

\noindent for $t =2, \dots, T$. For $t =1$, we assume a stationary distribution $h_{i,1} \sim \mathcal{N}\left( \mu_i, \frac{\sigma_{i}^{2}}{1-\phi_i^2}\right)$. The number of factors are defined using an upper bound as in \cite{aguilar2000bayesian}. However, in most practical applications, a small number of factors is sufficient to capture the dynamics of the covariance structure, as seen in \cite{bolfarine2024decoupling}; \cite{fruhwirth2024sparse}.

To address this, we impose a shrinkage prior on the columns of the $\mLambda$ matrix, which pushes irrelevant factors toward zero. For this purpose, we adopt the horseshoe prior proposed by \cite{carvalho2010horseshoe}. Conditional on the latent factors, this non-parametric VAR becomes $n$ unrelated regressions, estimated equation-by-equation.

\noindent Rewriting the model in terms of full-data matrices we have:
\begin{eqnarray}
\label{eq_by_eq}
    \mY = G(\mX) + \verror
\end{eqnarray}

\noindent where $\mX = \left(\vx_1, \dots, \vx_T\right)'$ and  $\mY = \left(\vy_1, \dots \vy_T \right)'$,  is a $T \times k$ and $T \times n$ matrix respectively. 

\subsection{The Learning Function: A BART Review}
\label{sec:bart}
Our choice estimating $\mG:\Real^{k}\rightarrow \Real^{n}$ is using a sequence of nonparametric decision tree ensemble such that:
\begin{eqnarray}
g_j(\mX) \approx \sum_{m=1}^M g_{j,m}\left(\mX \mid \cT_{j m}, \cM_{j m}\right), \quad \text{for } j=1,\dots, n. 
\end{eqnarray}

where $m$ is the number of trees in the ensemble. The general idea is to aggregate individuals ``weak learners" into a unified ``strong learner". Each regression tree $g_{j,m}(\cdot)$ defines a piecewise constant function based on the arrangement of split rules $\cT_{jm}$, associated with a $b_{jm}$-dimensional vector $\cM_{jm} = \left(\mu_{m,1}^{(j)}, \dots , \mu_{m,b_{jm}}^{(j)}\right)'$ of terminal nodes coefficients where $b_{jm}$ is the number of leaves per tree $m$ in equation $j$. 

There is established evidence in the literature that regression trees have demostrated strong empirical performance in a wide variety of contexts, including supervised learning (\cite{chen2016xgboost}; \cite{ke2017lightgbm}), casual inference (\cite{hahn2020bayesian}), density regression (\cite{orlandi2021density}). For more discussion see \cite{grinsztajn2022tree} and \cite{hill2020bayesian}.

The model prior structure follows \cite{chipman2010bart}. Conditioned on the model hyperparameters, BART priors are :

\begin{eqnarray}
    p\left((\cT_{j,1}, \cM_{,j1}), \dots, (\cT_{j,M}, \cM_{j,M}), \sigma_t{^2} \right) = p\left(\sigma_t^{2}\right)\prod_{m=1}^M p_{\cT}(\cT_{jk}) p_{\cM}(\cM_{jk} \mid \cT_{jk}).
\end{eqnarray}
 The prior distribution for the trees $p_{\cT}$ consists of two components. First a prior on the shape of the tree $\cT$ and second a prior on the splitting rules $[x_{q} \leq C_{q}]$ for each branch node of the tree. The prior on the tree structure includes a probability that a node of depth $d$ is ``terminal" or does not split. Starting from a root, the node will split with probability probability $\gamma(1 + d)^{-\beta}$. If that is not the case, the root is a terminal node. This iterates until all nodes at certain depth are terminal. We follow the convention introduced by \cite{chipman2010bart} by taking $\gamma = 0.95$ and $\beta = 0.2$.

If a node is split, the split rule, defined by a variable and a cutpoint, is sampled as follows. A variable index $q \in 1, \dots, k$ is sampled according to $q \sim Categorical\left(\vs\right)$, such that $\vs$ is probability vector where $s_{q} = \frac{1}{k}$. Subsequently, a cutpoint $C_q$ is sampled uniformly on the observed range of values of variable $x_{q,t}$ at the current node. Finally, for each terminal node $b_{jm}$ in the tree, we draw a mean parameter $\mu_{m,b_{jm}}^{(j)} \sim \mathcal{N}\left(0, \sigma_{\mu}^{2}/ M\right)$. 

\section{Minnesota BART}
\label{sec:Minnesota BART}
Recent advancements in the non-parametric Vector Autoregression (VAR) literature, particularly the integration of Bayesian Additive Regression Trees (BART) to extend the traditional VAR framework, provide compelling evidence of improved forecasting performance, as demonstrated by \cite{huber2022inference}. However, these recent developments do not adequately address scenarios in which the true data-generating process (DGP) is sparse. Additionally, the original BART prior used in \cite{huber2022inference} does not account for the temporal dependency structure present in macroeconomic data. Our contributions to the literature involve addressing these shortcomings within the vector autoregression framework.

A well-documented finding in macroeconomic forecasting—from univariate models for GDP, inflation, and interest rates to multivariate formulations is that simple random-walk or near-unit-root forecasts often perform reasonably well, particularly over short-to medium-term horizons, as initially shown by \cite{nelson1982trends}. This persistence aligns with the near-unit-root behavior frequently observed in economic time series.

Therefore, constructing a prior that embeds a random-walk assumption—where, in the absence of data, the best initial estimate is that today’s value equals yesterday’s value plus some small drift or shock—provides a natural way to incorporate this empirical regularity into the model.

The Minnesota prior, introduced by \cite{litterman1980bayesian} and \cite{doan1984forecasting}, is a shrinkage prior specifically designed to mitigate the issue of overparameterization commonly encountered in large linear VAR models.
The design for this prior is based on empirical evidence on macroeconomic time-series behavior, which suggests that variables often exhibit persistent dynamics that can be effectively captured through structured shrinkage.
It integrates several reasonable assumptions, including cross-variable shrinkage, where coefficients for lags of different variables are reduced more significantly than those for the variable's own lags. Additionally, it reflects the belief that higher-order lags contribute less to forecasting accuracy. For further details on the Minnesota prior, see \cite{kadiyala1997numerical}, \cite{karlsson2013forecasting}, and \cite{chan2020large}.

While the traditional BART framework has shown promise in time series analysis, It employs a uniform prior on splitting variables when sampling split rules—an assumption that is often unrealistic in multivariate forecasting settings. To address this limitation, we leverage the DART prior framework, presented by \cite{linero2018high}, to incorporate time dependency into our trees, drawing on insights from prior specification in the linear Vector Autoregression (VAR) literature.

By integrating the principles of the Minnesota prior into this framework, our approach introduces structured shrinkage that respects the temporal dependence inherent in macroeconomic data. This allows the estimation of large dynamic systems within a multivariate BART framework while preserving interpretability and forecasting accuracy.

For equation $n$, the prior for the splits probability  is defined:

\begin{eqnarray}
    \left(s_{1n}, \ldots, s_{kn}\right) \sim \operatorname{Dirichlet}(\phi_{1n}, \ldots, \phi_{kn})
\end{eqnarray}

The scale parameters of the Dirichlet distribution are defined are defined as follows:

$$
\phi_{in} =
\begin{cases}
\frac{\lambda_1}{l^2}, 
 & \text{for the scale on the $l$-th lag of variable $i$,}\\[6pt]
\frac{\lambda_2\cdot\rho}{l^2} , 
 & \text{for the coefficient on the $l$-th lag of variable $j$, $j \neq i$,}\\[6pt]
\end{cases}
$$

where $\rho = \frac{\sigma_{i}^{2}}{\sigma_{j}^{2}}$ represents the variance ratio of an AR($p$) process for each variable. This formulation induces a smooth shrinkage effect on the prior probabilities of splits. The choice of $ \lambda_j$, where $j =1,2$, determines the rate at which these probabilities decay. For our analysis, we set the hyperparameters to fixed values, specifically \( \lambda_1 = 1 \) and \( \lambda_2 = 0.5 \). While \( \lambda_1 \) and \( \lambda_2 \) can be estimated directly from the data, we choose to fix them to ensure analytical tractability and simplify the estimation process. A detailed discussion on prior elicitation is provided in Section \ref{prior_elic}.

However, this configuration does not lead to a sparse solution. It is possible to elicit a prior that explicitly favors a more parsimonious model with a smaller number of covariates. Conditional on the tree topology, and for a fixed $\lambda$, substituting $l^2$ with $k$, we obtain:

$$
\left(s_{1n}, \ldots, s_{kn}\right) \sim \operatorname{Dirichlet}\left(\frac{\lambda}{k}, \ldots, \frac{\lambda}{k}\right)
$$

The parameter \( \lambda \) governs the degree of sparsity introduced in the tree function. As demonstrated by \cite{linero2018high}, when both the number of predictors (\( k \)) and the number of branches (\( B \)) in the ensemble are large, the prior distribution of the number of relevant predictors (\( Q - 1 \)) can be approximated by a Poisson distribution with parameter \( \theta_B \), where \( \theta_B = \lambda \sum_{i=0}^{B-1} (\lambda + i)^{-1}\). The level of sparsity can be controlled by setting \( \lambda \) to a predetermined value. Under a fully Bayesian parameter selection framework, we follow the suggested approach of assigning \( \lambda \) a hyperprior, specifically \( \lambda/(\lambda + k)\sim \text{Beta}(0.5,1) \).

\begin{figure}[!h]
    \centering
     \includegraphics[scale=0.4]{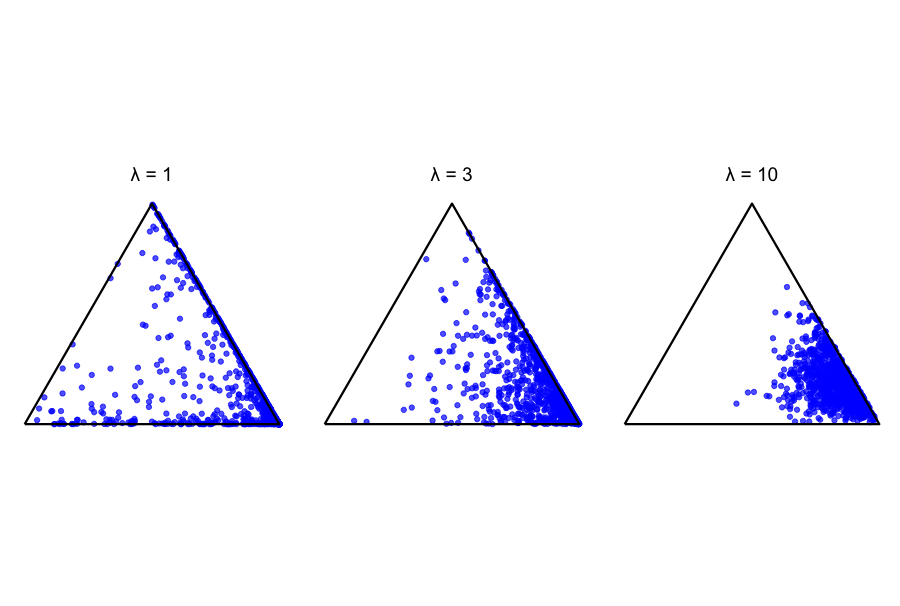}
    \caption{\textbf{Draws from  $\textit{Dirichlet}\left(\lambda, \frac{\lambda}{4},\frac{\lambda}{9} \right)$}. This figure illustrates the effect of varying $\lambda$ on the concentration parameters of the \textit{Dirichlet} prior on the simplex for $\lambda = \left(1, 3, 10\right)$. The vertices of the simplex correspond to one-sparse probability vectors, the edges represent two-sparse vectors, and the interior points indicate denser probability distributions. }
    \label{simplex_sim}
\end{figure}

Figure \ref{simplex_sim} illustrates how the value of $\lambda$  corresponds to different levels of sparsity. This approach may introduce some rigidity, but it aligns more closely with our objective of transparently examining the trade-off between sparsity across variables and to a variable's own lags in tree models. In Section \ref{prior_elic}, we discuss in detail how $\lambda$ affect an out-of-sample exercise. 

\section{Posterior Sampling Algorithm}
\label{sec:posterior sampling}

The model is estimated using a combination of traditional Bayesian inference techniques commonly employed in the VAR and BART literature. Tree sampling updates are performed using a Metropolis-Hastings (MH) algorithm, as proposed by \cite{chipman2010bart}, while most of the remaining steps leverage closed-form Gibbs Update. The conditional posteriors for the factor loadings and factors follow well-known Gaussian distributions. For the stochastic volatility components, present in both the factors and idiosyncratic innovations, we employ the efficient sampler outlined in \cite{kastner2014ancillarity}. In cases where a homoskedasticity assumption is applied, the model uses the traditional inverse-gamma prior.

As mentioned previously, conditioning on the covariance structure we can estimate the VAR equation-by-equation. The model can be rewritten the as a system of $n$ independent equations. Let $\mY_{\bullet j}$  denote the $j\text{-th}$ column of the matrix $Y$ as defined in \eqref{eq_by_eq}. Therefore we can write our dynamic system as :

\begin{eqnarray}
    \mY_{\bullet j} = \mG_{j}\left(\mX \right) + \mF \mLambda_{\bullet j}' + \veta_{\bullet j}
\end{eqnarray}

\noindent where $\mLambda_{\bullet j}'$ is the $j\text{-th}$ column of the factor loading matrix. This formulation reveals that our model is a generalized additive model, where the forest component approximates the relationship of $\vy_t$ with its lags, while a shared linear component across all equations captures the relationships among the variables.

\subsection{Sampling the Tree Structure}

The two main departures of the sampling strategy proposed by \cite{chipman2010bart} is regarding the partial residuals definition, that needs to take into account the factors and loadings structure and an additional step to update the vector of split probabilities $\vs$. To sample the trees using Bayesian backfitting as in the likelihood function depends of $\left(\cT_{jm}, \cM_{jm}\right)$ through the partial residuals that should be defined for our case as:

\begin{eqnarray}
    \mR_{jm}  \equiv  \mY_{\bullet j} - \mF \mLambda_{\bullet j}' - \sum_{m\neq m*}^{M}g_{jm}\left(\mX | \cT_{jm}, \cM_{jm}\right)
\end{eqnarray}

Therefore we can sample tree structure marginalizing over $\cM_{jm}$, such that:

\begin{eqnarray}
    p\left( \cT_{jm} |\mR_{jm}, \sigma_{j,t} \right) \propto p\left( \cT_{jm}\right) \int p\left( \mR_{jm} | \cM_{jm}, \cT_{jm}, \sigma_{j,t}\right) p\left( \cM_{jm}|\cT_{jm}, \sigma_{j,t}\right)d\cM_{jm}
\end{eqnarray}
\noindent can be obtained in a closed form solution up to a constant. Allowing to carry out each draw from $\left(\cT_{jm}, \cM_{jm} |\mR_{jm}, \sigma_{j,t}\right)$ sequentially. 

To draw the probability split vector $\vs$, we follow \cite{linero2018high} and leverage the conjugacy between the Dirichlet prior and multinomial sampling, enabling a full-conditional Gibbs update given by:

\begin{eqnarray}
    \left(s_1, \dots, s_k\right) \sim \text{Dirichlet}\left( \phi + m_1, \dots, \phi + m_k \right),
\end{eqnarray}
    
\noindent where $\phi$ represents the shape parameter of the Dirichlet and Minnesota specification, and $m_k$ denotes the number of splitting rules for predictor $k$ in the ensemble.

\subsection{Sampling the Loadings and Factors}
The factor loadings $\mLambda$ are drawn from a full conditional distribution that follows a Gaussian distribution in a standard form. For each row of  $\mLambda$, denoted as $\mLambda_i$, we sample as follows:

\begin{eqnarray}
\mLambda_i | \bullet \sim \mathcal{N}(\bar{\mL}_i, \bar{\mW}_i),    
\end{eqnarray}

\[
\bar{\mW}_i = \left(\mF_i^\prime \mF_i +\mW_i^{-1}\right)^{-1},
\]

\[
\bar{\mL}_i = \bar{\mW}_i \left(\mF_i^\prime \tilde{y}_i\right).
\]

Here, $\mF_i$ is the $t$-th row $\vf_t / e^{h_it / 2}$, and the $t$-th element of $\tilde{y}_i$ is given by $\left(y_{it} - f_i(x_t) \right) / e^{h_it / 2}$. The matrix $\mW_i$ is a prior variance-covariance matrix of dimension $(n r) \times (n r)$. Since the number of factors are determined by an upper bound, we sample from a horseshoe prior using the auxiliary sampler proposed in \cite{makalic2015simple} for each column of $\mLambda$. The factors are generated on a  $t\text{-by-}t $ basis using Gaussian distributions as in \cite{aguilar2000bayesian}.

\section{Forecasting Macroeconomic Variables}

We conduct a forecasting exercise using multivariate Bayesian additive regression trees to compare our proposed sparse and Minnesota-BART priors with the baseline BART prior structure from \cite{huber2022inference} 

\subsection{Data}

We use a dataset consisting of 22 U.S. quarterly variables covering the period from 1965Q1 to 2019Q4. The data is sourced from the FRED-QD database at the Federal Reserve Bank of St. Louis, as described in \cite{mccracken2016fred}. The dataset includes a range of standard macroeconomic and financial variables, such as real GDP, industrial production, inflation rates, labor market indicators, and interest rates. These variables are transformed to achieve stationarity, typically by computing growth rates. We will include 13 lags of the endogenous variables in our model. A detailed description of the variables and their transformations can be found in Appendix \ref{sec:data}.

\subsection{Predictive Distribution}

We begin by presenting predictive evidence of the effectiveness of our proposed priors and their impact on predictive accuracy. To achieve this, we construct a recursive forecasting design, using 1965Q1 to 2004Q4 as the initial training period. We employ an expanding window strategy: after performing $h$-step-ahead forecasts, we incorporate the next observation into the dataset and re-estimate the model, obtaining a new draw from the predictive density with the updated information. This process is repeated iteratively until all available data has been utilized. The one-step-ahead predictive distribution is given by 

\begin{eqnarray}
    p\left(\vy_{t+1} | \vy^{t}\right) = \int p\left( \vy_{t+1} | \vy^{t}, \thmod \right)  p\left(\thmod | \vy^{t}\right) d\thmod,
\end{eqnarray}

where $\vy^{t}$ represents the historical time series up to time $t$, i.e., $\vy^{t} = \left(y_1, \dots, y_{t}\right)$. The parameter $\thmod$ encapsulates all unknown parameters of the model. This integral is solved using the standard Monte Carlo approach:

$$
p\left(\vy_{t+1} | \vy^{t}\right)  \ \approx \, \frac{1}{M} \sum^M_{m=1}  p(\vy_{t+1}|\vy^{t},\thmod^{(m)}),
$$

where the one-step-ahead predictive density $p\left(\vy_{t+1} | \vy^{t}, \thmod \right)$ is Gaussian, conditional on knowing $\thmod$. Therefore, for each draw $\thmod^{(m)}$ from the posterior distribution $p\left(\thmod | \vy^{t}\right)$, we obtain the predictive density:

$$
\vy_{t+1}|\vy^{t}, \thmod^{(m)} \sim \mathcal{N}\left(\mF^{(m)} \left(\mX_{t+1}\right), \Sigma_{t+1}^{(m)}\right),
$$

where $\mF^{(m)}(\cdot)$ is generated by our tree sampling algorithm, and $\Sigma_{t+1}^{(m)}$ is drawn using the covariance structure specified in Eq. \ref{cov}. For the stochastic volatility (SV) specification, as outlined in Eq. \ref{sv}, the forecasts of $\Omega_{t+1}$ and $H_{t+1}$ are obtained as follows. Given the posterior draws of $h_{it}^{(m)}$, we simulate $h_{it+1}^{(m)}$ from a conditional normal distribution with mean $\mu^{(m)}_{i}+\phi^{(m)}_{i}(h^{(m)}_{i,t} -\mu^{(m)}_{i})$ and variance $\sigma^{2(m)}_{h}$. Higher-order forecasts are computed recursively. In the following section, note that $\vy_{t+1}$ corresponds to a one-quarter-ahead prediction, i.e., three months. We use the Gibbs sampler described in the section \ref{sec:posterior sampling} to obtain 5,000 posterior, after a burn-in period of 30,000.

\subsection{A point forecast comparison of the different priors}

\begin{figure}[ht!]
\centering
\includegraphics[width=\textwidth,height=10cm]{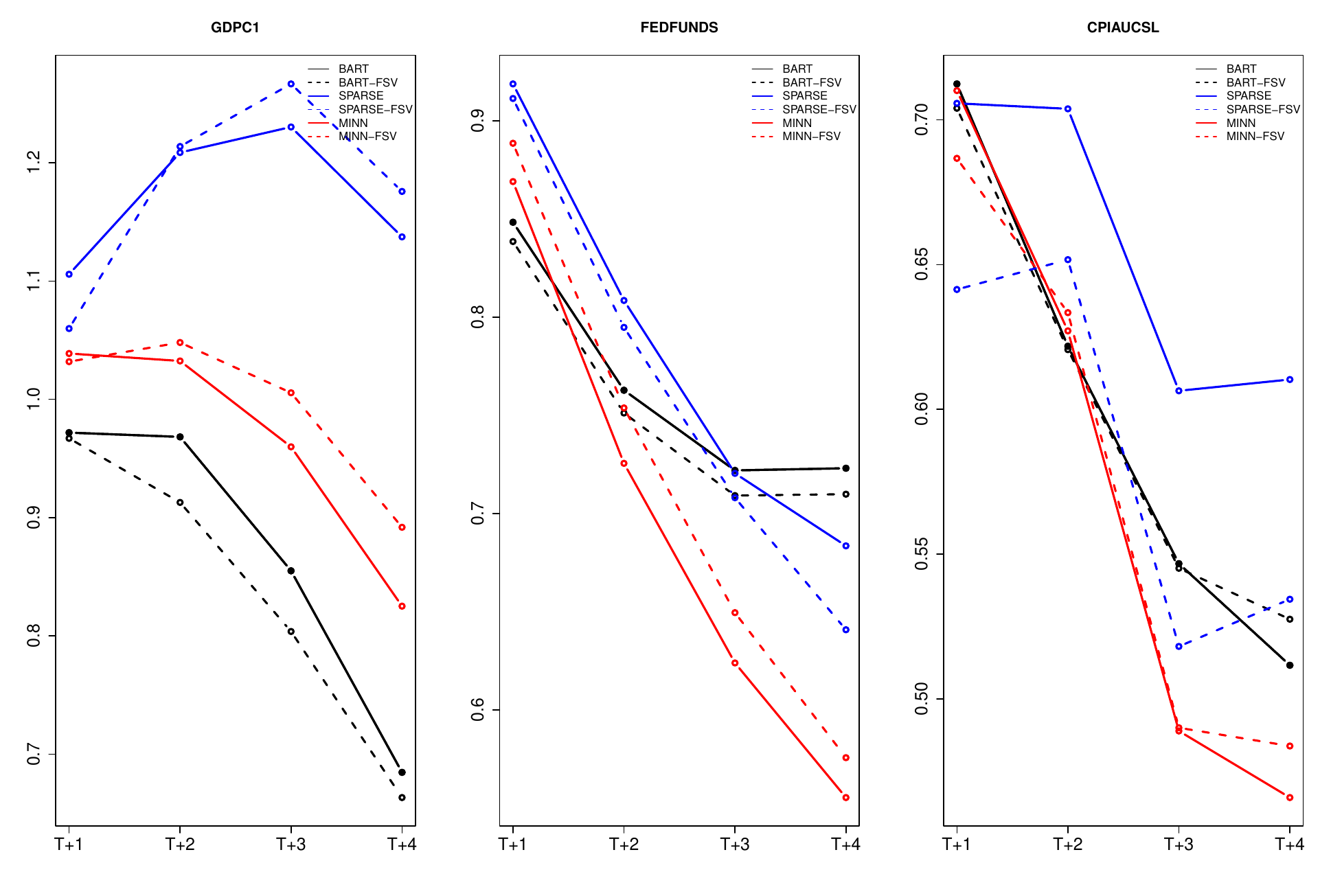}
\caption{\textbf{Point Forecast Comparison.} This figure reports the Relative RMSE of the variables of interest compared to the baseline BVAR-SV, using the Minnesota-BART prior, the Sparse Prior, and the BART prior. A value below one indicates that the model outperforms the benchmark, while a value above one suggests weaker performance. Each probability split prior specification for the mean function is shown under both the homoskedastic and stochastic volatility (SV) settings, where the former is represented by a continuous line and the latter by a dashed line.  
}
\label{fig:rmse}
\end{figure}

We use a standard small-scale Minnesota BVAR with stochastic volatility (SV) as a benchmark to evaluate our model's performance across three key variables: real GDP growth (GDPC1), inflation (CPIAUCSL), and the Federal Funds Rate (FEDFUNDS). To facilitate this comparison, we compute the (relative) root mean square prediction error (RMSPE). A value below one indicates that the model outperforms the benchmark, while a value above one suggests weaker performance. Importantly, we account for heteroskedasticity in each prior specification. A priori, we expect that the combination of our proposed priors and time variation in volatilities will enhance point forecasts. The results are summarized in Figure \ref{fig:rmse}.

Our findings reveal a consistent pattern: predictive accuracy improves as the forecasting horizon extends. Compared to the linear baseline, incorporating a richer information set and a non-linear approach enhances point forecasts. The SV variants generally improve precision relative to their homoskedastic counterparts. However, the numerical differences are often small, particularly when comparing sparse and non-sparse versions of the prior. Importantly, for inflation forecasting, the introduction of stochastic volatility (SV) leads to a more pronounced improvement in predictive accuracy for the sparse specification compared to its non-SV counterpart. Regarding prior choice, with the exception of real GDP growth, a smoother shrinkage approach in the variable splits tends to outperform its sparse alternative for the variables of interest. This suggests that, for certain variables, a prior that imposes gradual shrinkage—leveraging more information compared to a sparse prior—is preferable.

\subsection{Comparing the priors through log predictive density scores}
Although point forecast exercises are important, they provide only a partial assessment of model performance. To obtain a more comprehensive evaluation, it is essential to consider additional metrics that account for the model's ability to predict higher-order moments of the predictive distribution for the variable of interest. Therefore, is necessary to utilize a metric to access the accuracy of density forecasts. As discussed by \cite{geweke2010comparing}, log predictive density scores (LPDS) are often used to compare different models. In this paper, we use the LPDS as a metric to evaluate and compare the performance of various BART prior specifications. We will make slightly change of notation when presenting this calculations. The first $t_0$ time series observations, $\vy^{tr} = (y_1, \ldots, y_{t_0})$, are designated as the ``training sample," while the remaining observations, $y_{t_0+1}, \ldots, y_T$, are used for evaluation based on the log predictive density:

\begin{eqnarray} \label{marlikr}
\LPS = \log p(y_{t_0+1}, \ldots, y_T \mid \vy^{tr}) = 
\sum_{t=t_0+1}^T \log p(\vy_t \mid \vy^{t-1})
\end{eqnarray}

In Equation (\ref{marlikr}), $p(\vy_t \mid \vy^{t-1})$ represents the one-step-ahead predictive density for time $t$. This density is evaluated at the observed value $\vy_t$. Note that this framework not only works when evaluating the joint performance for higher order moments of the predictive distribution for the multivariate model, but also the marginal density scores, i.e, we have the LPDS for the three variables of interest as well. 

\begin{figure}[ht!]%
\centering
 \includegraphics[scale=0.8]{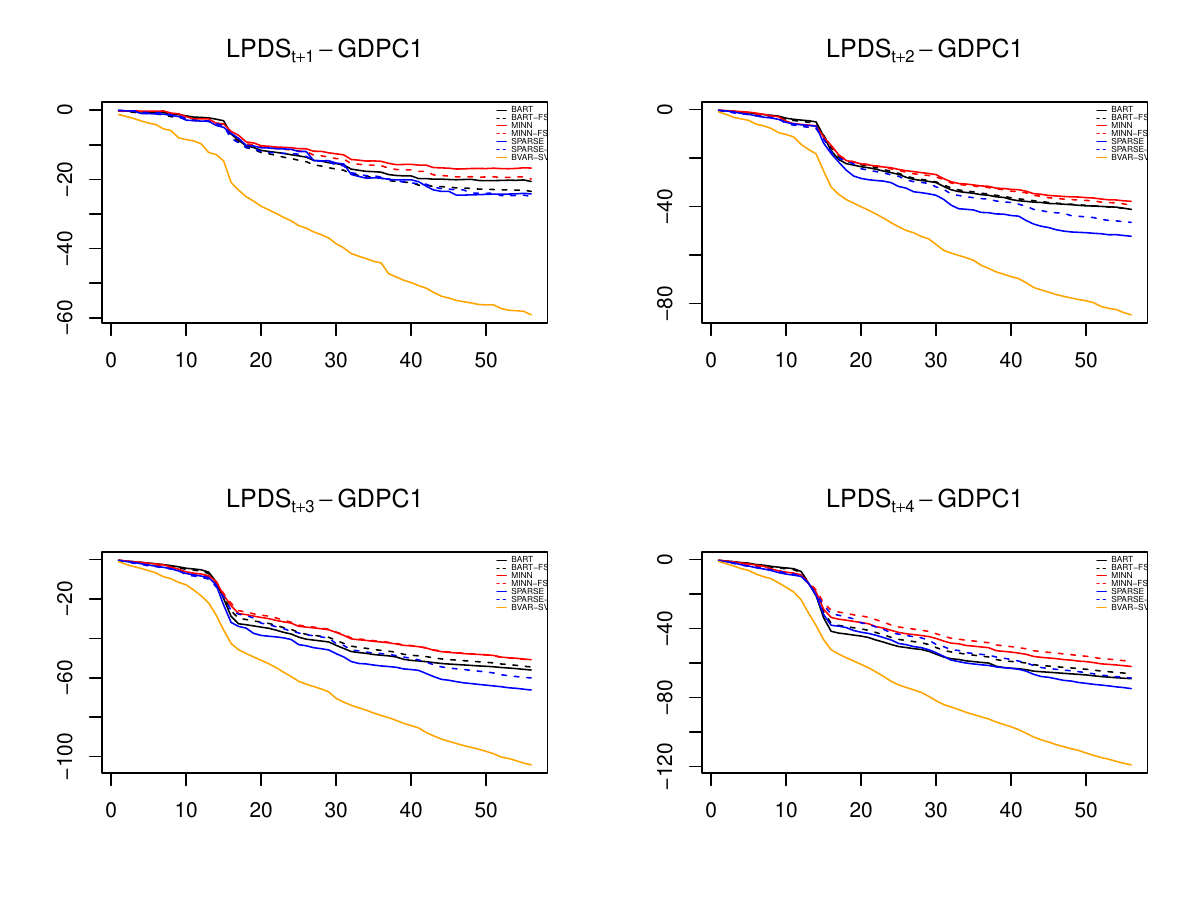}
\caption{\textbf{Marginal Log Predictive Density Score.} This figure reports the Marginal Log Predictive Density Score (LPDS) for GDPC1. Cumulative Marginal log predictive scores for the last 56  time point (labeled with time index $T-t_0$, where $t_0=160$). Each probability split prior specification for the mean function is shown under both the homoskedastic and stochastic volatility (SV) settings, where the former is represented by a continuous line and the latter by a dashed line.} 
\label{fig:lpds_uni_gdp}
\end{figure}

Figure \ref{fig:lpds_uni_gdp} shows that while our prior does not perform as well in point forecasting for GDP, it excels in predictive distribution forecasting. The Minnesota prior achieves the best performance, with its FSV specification closely following. Notably, this prior alternative not only outperforms the basic BART prior but also surpasses its simpler linear counterpart.

\begin{figure}[ht!]%
  \centering
 \begin{tabular}{c}
\includegraphics[scale=0.8]{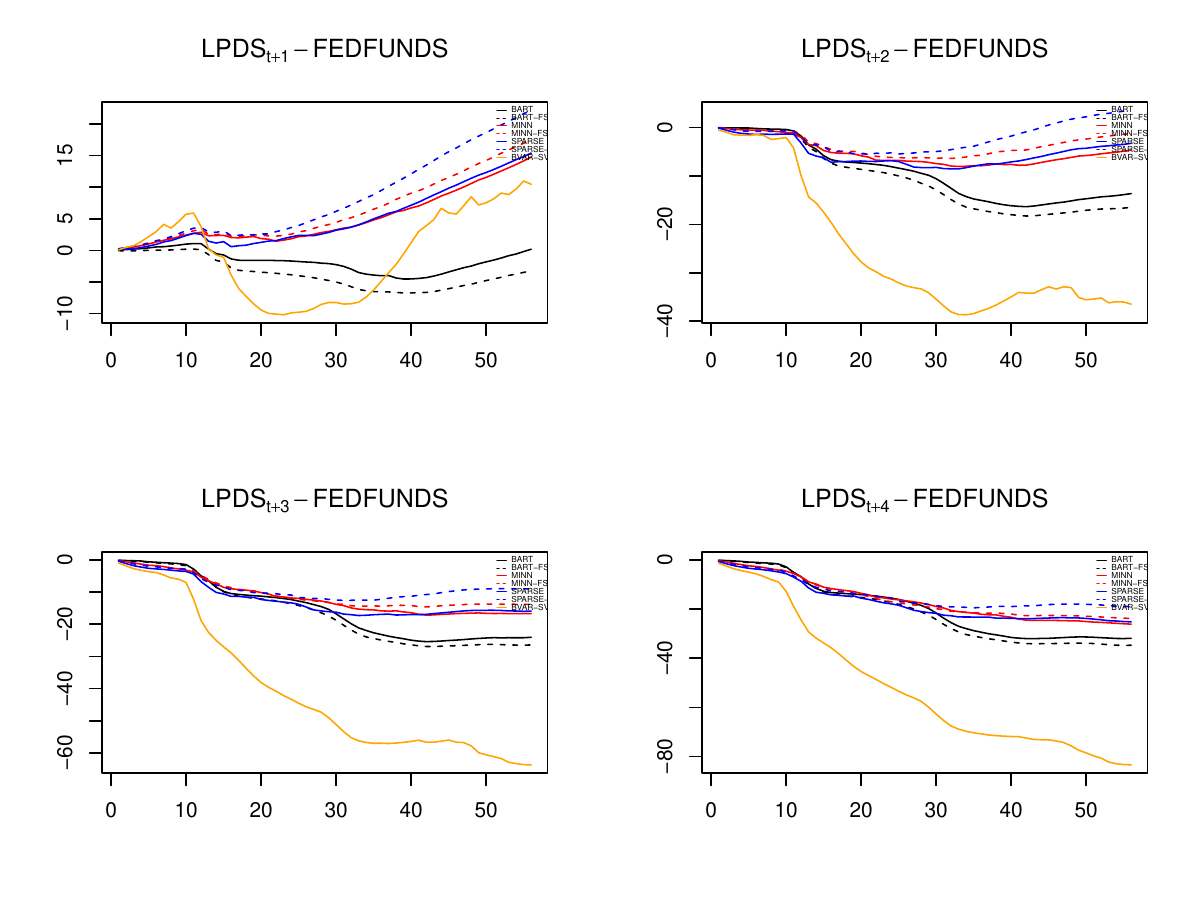}
\end{tabular}
\caption{\textbf{Marginal Log Predictive Density Score.} This figure reports Marginal Log Predictive Density Score (LPDS) for FEDFUNDS. Cumulative Marginal log predictive scores for the last 56  time point (labeled with time index $T-t_0$, where $t_0=160$). Each probability split prior specification for the mean function is shown under both the homoskedastic and stochastic volatility (SV) settings, where the former is represented by a continuous line and the latter by a dashed line.} 
\label{fig:lpds_uni_ff}
\end{figure}

For the FEDFUNDS variable, the results follow a different pattern. As shown in Figure \ref{fig:lpds_uni_ff}, the more sparse specification underperforms in point forecasting when compared to the linear benchmark. However, it dominates across all horizons in predictive distribution forecasting. The smooth shrinkage prior emerges as the second-best choice.

\begin{figure}[!htpb]%
\centering
\includegraphics[scale=0.8]{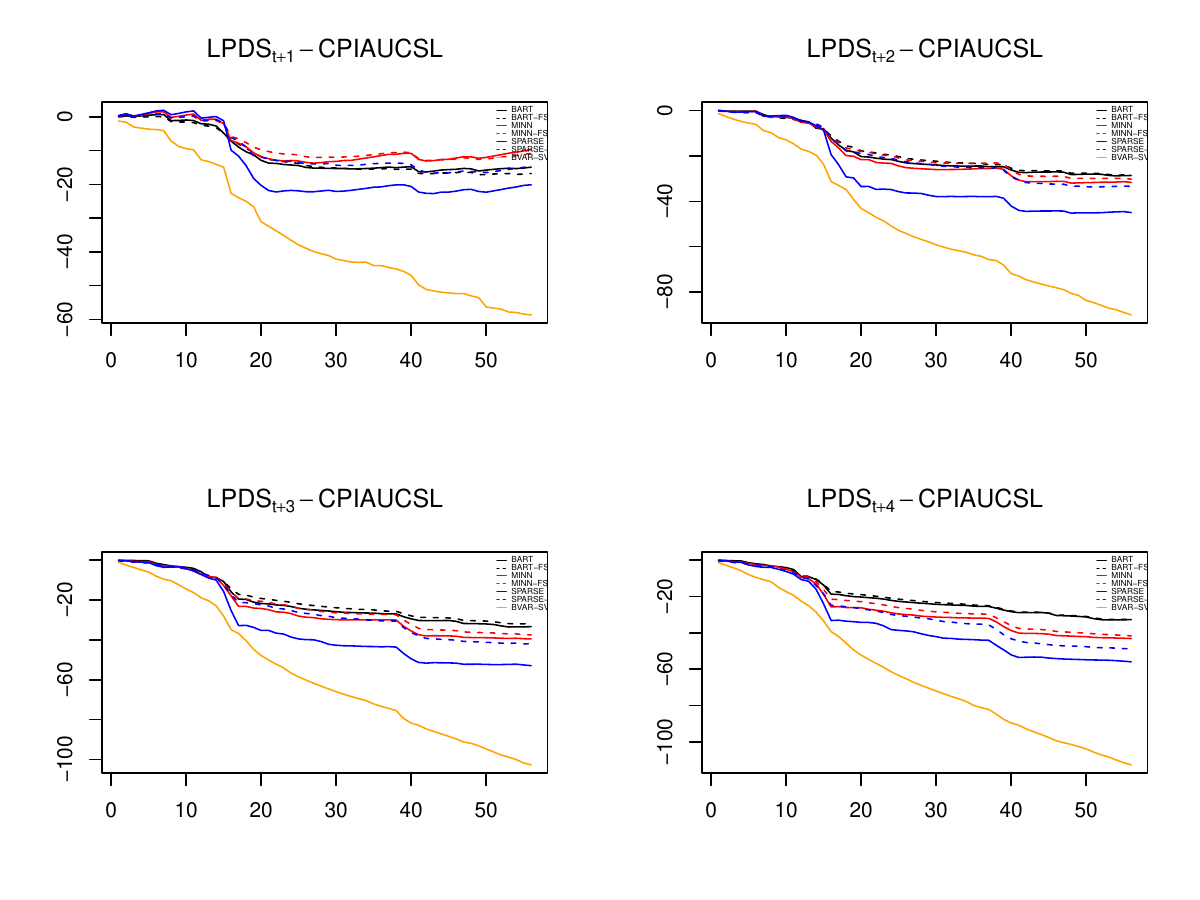}
\caption{\textbf{Marginal Log Predictive Density Score. }This figure reports the Marginal Log Predictive Density Score (LPDS) for CPIAUCSL. Cumulative Marginal log predictive scores for the last 56  time point (labeled with time index $T-t_0$, where $t_0=160$). Each probability split prior specification for the mean function is shown under both the homoskedastic and stochastic volatility (SV) settings, where the former is represented by a continuous line and the latter by a dashed line.} 
\label{fig:lpds_uni_cpi}
\end{figure}

For the CPI variable, the results favor the baseline prior with an SV correction, except for the first forecasting horizon, as shown in Figure \ref{fig:lpds_uni_cpi}. Even in this case, our smooth shrinkage option remains a close second in forecasting the predictive density distribution.

\begin{figure}[!htpb]%
\centering
\includegraphics[scale=0.5]{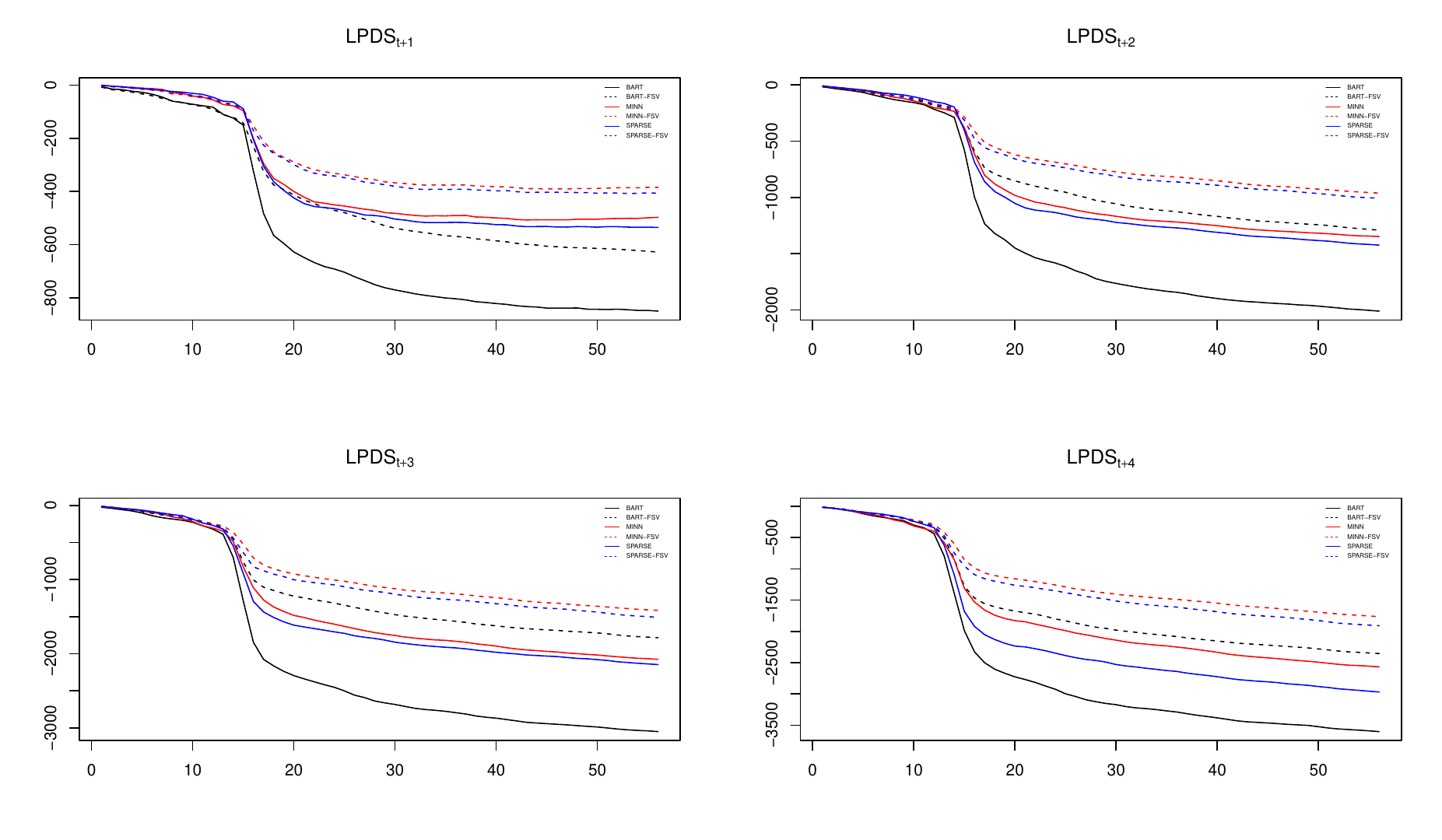}
\caption{\textbf{Cumulative Log Predictive Density Scores.} Cumulative log predictive scores for the last 56  time point (labeled with time index $T-t_0$, where $t_0=160$). Each probability split prior specification for the mean function is shown under both the homoskedastic and stochastic volatility (SV) settings, where the former is represented by a continuous line and the latter by a dashed line.} 
\label{fig:forecasting}
\end{figure}

This results suggest that, for both point forecasts and marginal density, the alternative prior structures proposed in this paper are competitive and, in some cases, even outperform the baseline BART model. Figure \ref{fig:forecasting} shows that when evaluating the (joint) predictive density, our alternatives, which incorporate varying levels of shrinkage, consistently outperform the basic prior, particularly when heteroskedasticity is accounted for, as initially hypothesized. Notably, the Minnesota specification outperforms the sparse alternative across all time horizons. 

\subsection{In sample features of our model}

\begin{figure}[htpb!]%
\centering
\includegraphics[width=15cm,height=10cm]{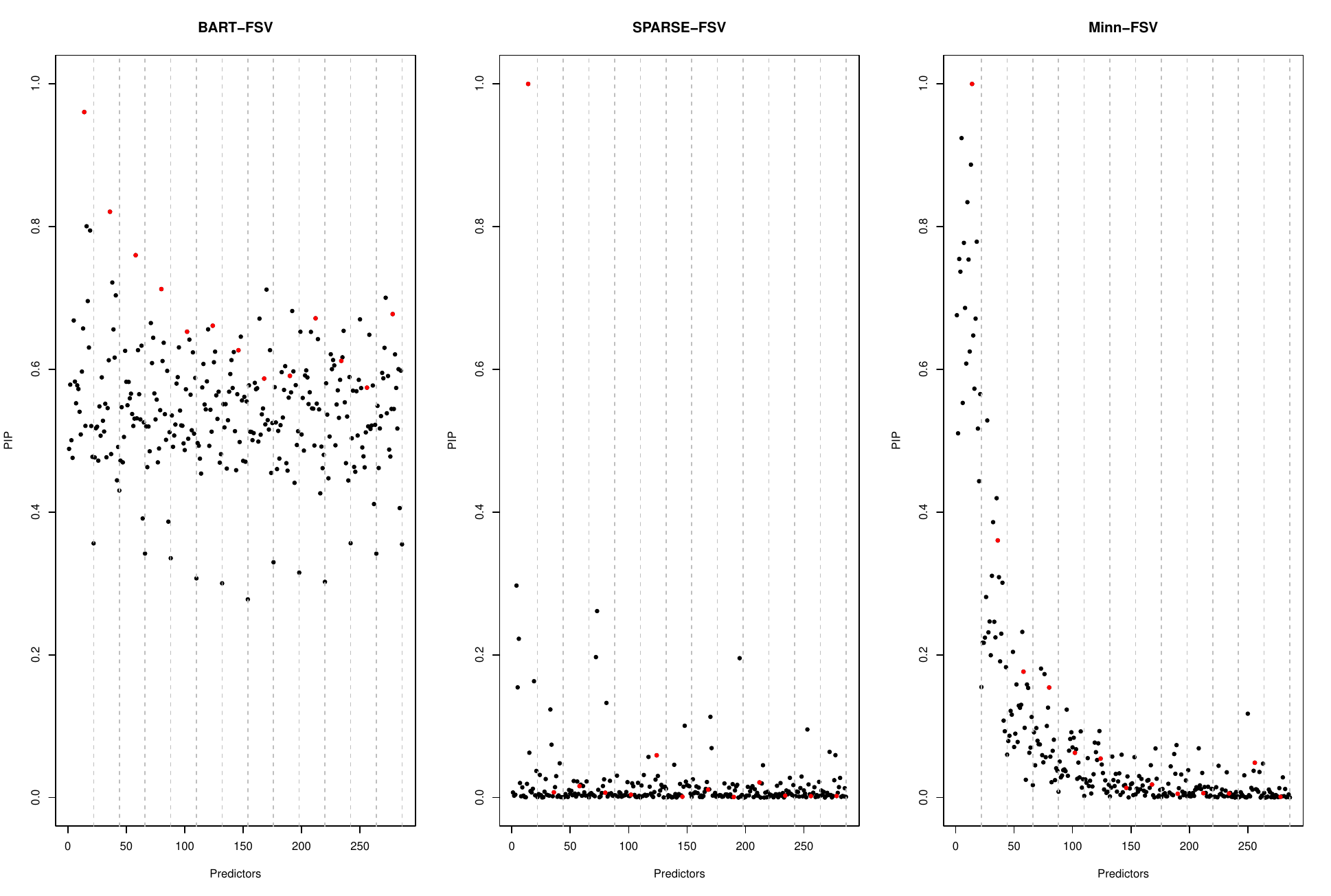}
\caption{\textbf{Posterior Inclusion Probability. }In-sample Posterior Inclusion Probability (PIP) results for the CPI variable, before expanding window exercise. The dashed vertical lines indicate the bin boundaries corresponding to different lag orders. The highlighted red dots represent the variable's own lags.}
\label{fig:pip}
\end{figure}

Our prior modifications introduce key advantages to multivariate BART analysis. In the baseline BART framework, variable importance is typically assessed by counting the number of times each feature appears in a splitting rule. However, this approach has limitations. Since the traditional prior imposes a uniform split probability across all features, variable importance can only be inferred if there is a trade-off between the number of trees and interpretability. In other words, improving interpretability often comes at the expense of predictive performance \cite{chipman2010bart}; \cite{bleich2014bayesian}; \cite{linero2018high}. This limitation does not apply to our proposed prior structures. To illustrate, Figure \ref{fig:pip} presents the Posterior Inclusion Probability (PIP) for each prior choice. The PIP is defined as:
$$
\text{PIP}_j = P(\text{predictor } j \text{ appears in the ensemble} | \text{Data})
$$

We present the PIP for the in-sample results of our model before applying the expanding window, focusing on the Inflation variable. As shown in Figure \ref{fig:pip}, the BART prior specification distributes splits relatively evenly across all features in the model. In contrast, the sparse prior specification shrinks the split probability of most variables to near zero while preserving the variable’s own first lag as the most significant predictor. This aligns with the Bayesian VAR literature, which finds that the AR(1) term explains the largest share of variation in the variable of interest. For the Minnesota prior, the expected decay in PIP follows a structured lag hierarchy, preserving the anticipated importance ordering among the split variables.

\section{Prior Elicitation}
\label{prior_elic}

As previously discussed, the choice of \( \lambda \) is of critical importance, as it plays a central role in determining the expected level of shrinkage in the model. One approach we have presented is to select \( \lambda \) based on a shrinkage target informed by the econometric literature and subject-matter considerations. To further investigate its impact, we analyze different levels of \( \lambda \), specifically considering a grid of values: \( \lambda_1 = \{1, 3, 5, 10, 20\} \) and \( \lambda_2 = \{0.5, 1, 1.5, 2.5, 5, 10\} \). We then examine its effects on the log-predictive density score in comparison to the standard BART prior.

\begin{figure}[!htpb]
  \centering
  \includegraphics[scale = 0.8]{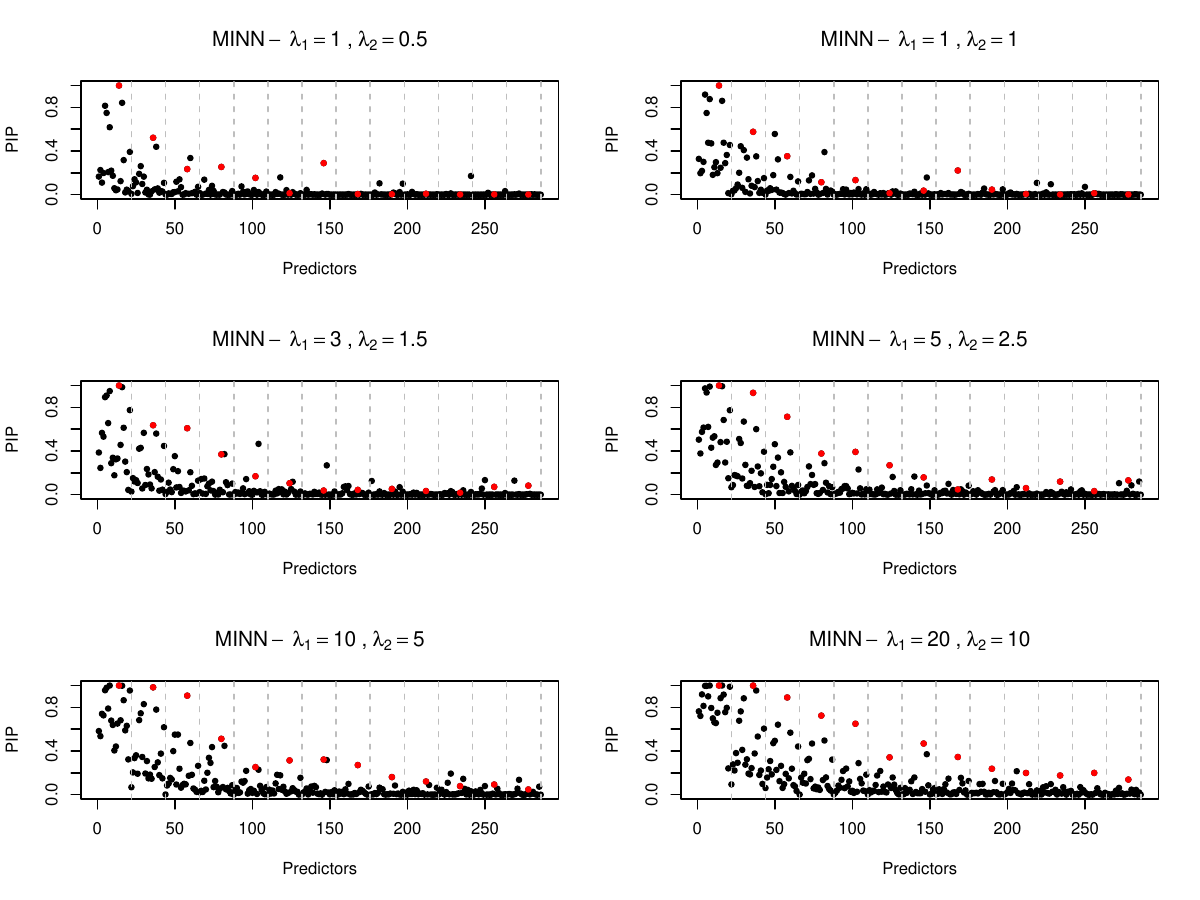}
  \caption{\textbf{Posterior Inclusion Probability for different shrinkage parameters. }In-sample Posterior Inclusion Probability (PIP) results for the CPI variable for each different $\lambda_i$ combination. The dashed vertical lines indicate the bin boundaries corresponding to different lag orders. The highlighted red dots represent the variable's own lags. }
  \label{fig:pip_lambdas}
\end{figure}

First, we examine how the choice of the smooth shrinkage parameter influences the ``hyperbolic" shape of the posterior inclusion probability (PIP) for the inflation variable in the in-sample analysis. As shown in Figure \ref{fig:pip_lambdas}, an increase in \( \lambda \) values results in a slower decay of PIP, reflecting that the shrinkage of both lags and cross-lags does not reach zero as quickly as with our reference values of \( \lambda_1 = 1 \) and \( \lambda_2 = 0.5 \). 

\begin{figure}[htpb]
  \centering
  \includegraphics[scale = 0.8]{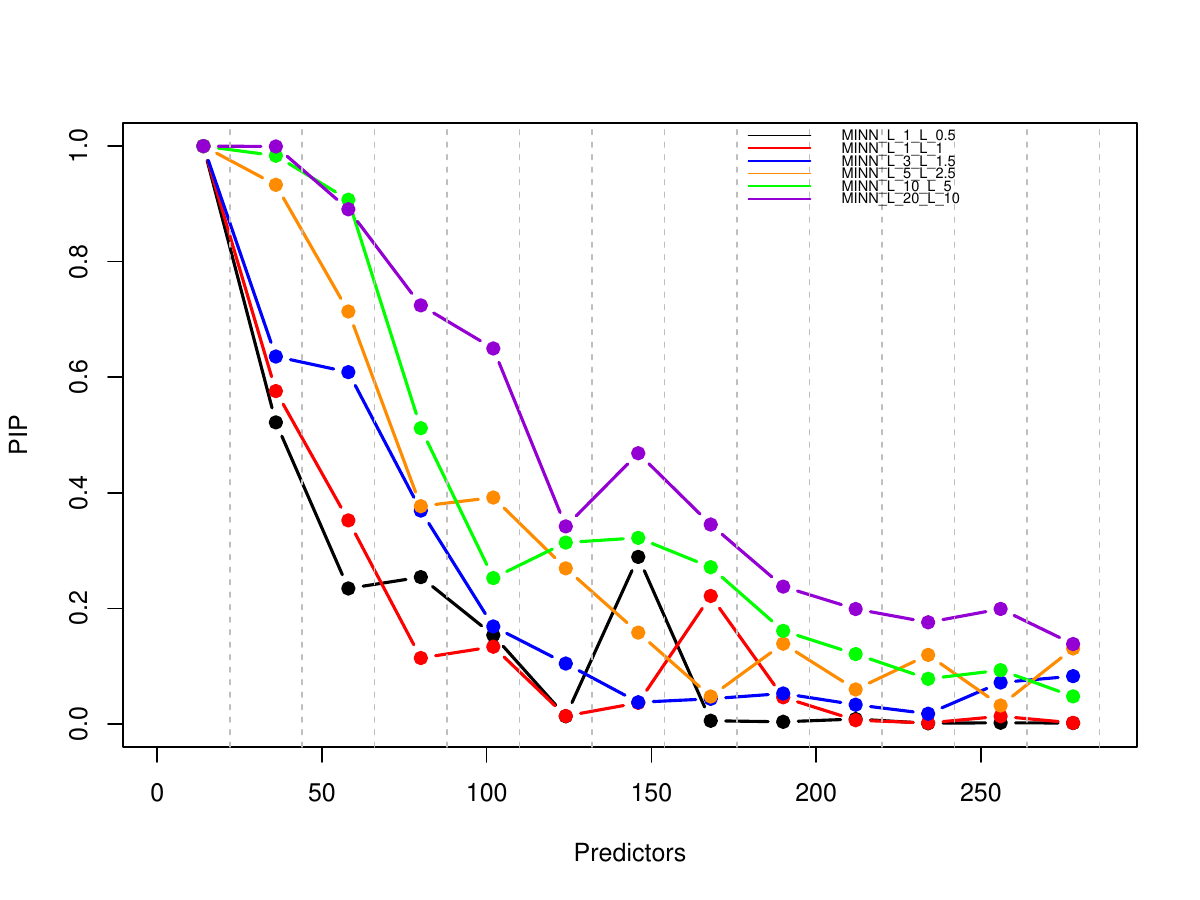}
  \caption{\textbf{Own-Lag Posterior Inclusion Probability}. In-sample Posterior Inclusion Probability (PIP) for the CPI's own lag across different grid values of $ \lambda_1 = \{1, 3, 5, 10, 20\}$  and $\lambda_2 = \{0.5, 1, 1.5, 2.5, 5, 10\}. $}
  \label{fig:pip_cpi_lambdas}
\end{figure}

\begin{figure}[htpb!]
  \centering
  \includegraphics[scale = 0.8]{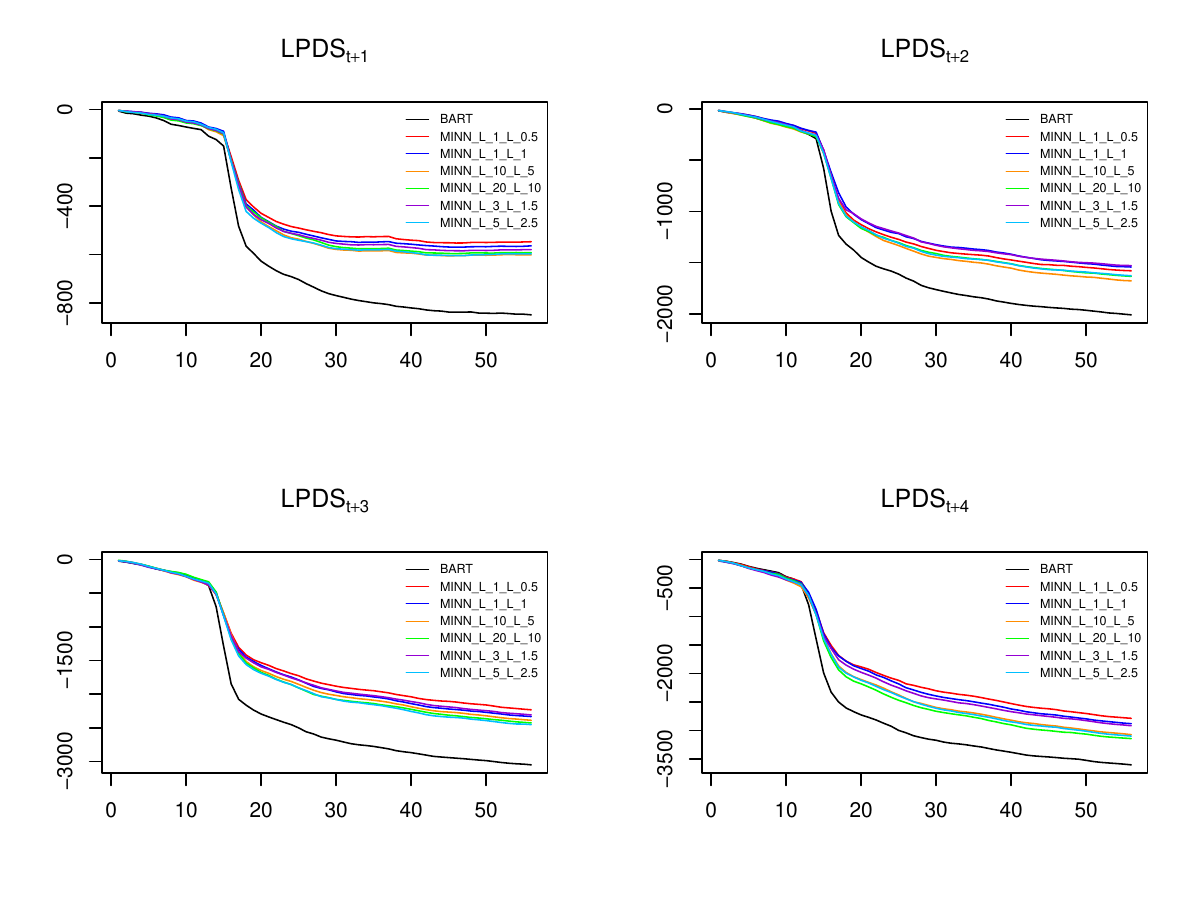}
  \caption{\textbf{Log Predictive Density Score for different shrinkage values. }Cumulative log predictive scores for the last 56 time points (labeled with time index $T-t_0$, where $t_0=160$), across different grid values of $\lambda_1 = \{1, 3, 5, 10, 20\}$  and $\lambda_2 = \{0.5, 1, 1.5, 2.5, 5, 10\}.$}
  \label{fig:lpds_lambdas}
\end{figure}

Additionally, when comparing different PIPs for CPI’s own lag, we observe the influence of the hyperparameter choice on the rate at which the inclusion probability decays, as illustrated in Figure \ref{fig:pip_cpi_lambdas}. To compare different prior configurations we will use the log predictive density score, calculated as how was presented in the sections before. Figure \ref{fig:lpds_lambdas} shows that besides any configuration that you choose for the prior, it is clear that for accurate density forecasts, its necessary to take into account time dependency when constructing the prior.

Overall, our analysis highlights the crucial role of \( \lambda \) in shaping the degree of shrinkage in the model and its impact on both variable selection and predictive performance. The results demonstrate that higher values of \( \lambda \) lead to a more gradual decay in posterior inclusion probabilities, preserving the influence of lags and cross-lags for a longer range. These findings reinforce the importance of carefully selecting the shrinkage parameter, as it directly affects model interpretability and forecasting accuracy.

\section{Concluding Remarks and Future Research}

The classic BART prior for split rules, which samples split variables with uniform probability, implicitly assumes that the mean function is not sparse or time-dependent in its input features, limiting inference on variable importance and hindering effective high-dimensional analysis. This paper introduces a framework that integrates insights from the literature on sparse priors for BART and Bayesian VARs. The proposed model allows for shrinkage in split probabilities, enabling the estimation of large dynamic systems within a multivariate BART framework, unlike existing approaches. Additionally, we demonstrate that incorporating a prior akin to the linear Minnesota prior introduces smooth shrinkage and time dependence information during prior elicitation.

Our approach was illustrated using a large U.S. dataset, where we showed that the proposed priors yield substantial improvements in forecast accuracy, particularly for higher-order moments. In this context, the Minnesota specification appears to be especially effective in extracting forecasting gains by introducing a smoother shrinkage approach compared to the sparse alternative. While point forecast accuracy varies across the variables of interest, nonlinear models often outperform linear specifications, their gains are substantial when they do, whereas losses to linear models are relatively small. Moreover, our prior structure demonstrates predictive gains for key macroeconomic variables, such as the Federal Funds Rate and inflation, highlighting its practical relevance for economic forecasting.

Although our model focuses on a reduced-form specification for forecasting, the framework can also be adapted for structural analysis using techniques such as Generalized Impulse Response Functions (GIRFs) (\cite{koop1996impulse}) or Local Projection (LP) estimation (\cite{jorda2005estimation}). Future work could explore alternative sampling methods or algorithmic optimizations to enhance scalability and reduce computational costs. Furthermore, expanding this approach to account for richer structural dynamics, such as incorporating time-varying parameters or state-dependent effects, could significantly enhance its applicability in macroeconomic modeling.

\pagebreak
\bibliographystyle{apalike}
\bibliography{references}

\pagebreak

\appendix
\section{Data description} \label{sec:data}
\begin{table}[!htbp]
{\tiny
\begin{center}
\caption{Data description.\label{tab:data}}
\scalebox{0.8}{
\begin{tabular}{lllccccc}
\toprule
\multicolumn{1}{l}{\ }&\multicolumn{1}{c}{\bfseries Mnemonic}&\multicolumn{1}{c}{\bfseries Description}&\multicolumn{1}{c}{\bfseries Trans.}&\multicolumn{1}{c}{\bfseries VAR-8}&\multicolumn{1}{c}{\bfseries VAR-22}\tabularnewline
\midrule
~~&GDPC1 (RGDP) &Real Gross Domestic Product&$2$&x&x\tabularnewline
~~&CE16OV (EMP) &Civilian Employment (Thousands of Persons)&$2$&x&x\tabularnewline
~~&AWHMAN (AWH) &Average Weekly Hours of Production and Nonsupervisory Employees:  Manufacturing&$1$&x&x\tabularnewline
~~&CPIAUCSL (CPI) &Consumer Price Index for All Urban Consumers:  All Items&$2$&x&x\tabularnewline
~~&CES3000000008x (AHE) &Real Average Hourly Earnings of Production and Nonsupervisory Employees: Manufacturing&$2$&x&x\tabularnewline
~~&INDPRO&IP:Total index Industrial Production Index (Index 2012=100)&$2$&x&x\tabularnewline
~~&FEDFUNDS (FFR) &Effective Federal Funds Rate (Percent)&$1$&x&x\tabularnewline
~~&S.P.500 (SP500) &S\&P's Common Stock Price Index:  Composite&$3$&x&x\tabularnewline
\midrule
~~&PCECC96&Real Personal Consumption Expenditures&$2$&&x\tabularnewline
~~&FPIx&Real private fixed investment &$2$&&x\tabularnewline
~~&UNRATE&Civilian Unemployment Rate (Percent)&$1$&&x\tabularnewline
~~&CES0600000007&Average Weekly Hours of Production and Nonsupervisory Employees:  Goods-Producing&$1$&&x\tabularnewline
~~&CLAIMSx&Initial Claims&$2$&&x\tabularnewline
~~&HOUST&Housing Starts: Total: New Privately Owned Housing Units Started&$2$&&x\tabularnewline
~~&CES0600000008&Average Hourly Earnings of Production and Nonsupervisory Employees:&$2$&&x\tabularnewline
~~&PAYEMS& Emp:Nonfarm All Employees: Total nonfarm (Thousands of Persons)&$2$&&x\tabularnewline
~~&CUMFNS&Capacity Utilization:  Manufacturing (SIC) (Percent of Capacity)&$1$&&x\tabularnewline
~~&PERMIT&New Private Housing Units Authorized by Building Permits&$2$&&x\tabularnewline
~~&BUSLOANSx&Real Commercial and Industrial Loans, All Commercial Banks&$2$&&x\tabularnewline
~~&BAA10YM&Moody's Seasoned Baa Corporate Bond Yield Relative to Yield on 10-Year Treasury&$1$&&x\tabularnewline
~~&GS10TB3Mx&10-Year Treasury Constant Maturity Minus 3-Month Treasury Bill, secondary market&$1$&&x\tabularnewline
~~&TB3SMFFM&3-Month Treasury Constant Maturity Minus Federal Funds Rate&$1$&&x\tabularnewline
\midrule
\bottomrule
\end{tabular}}
\begin{minipage}{1.09\textwidth}
\vspace*{5pt}
\scriptsize 
\noindent \textit{Notes:} The data used is the quarterly version of the dataset proposed in \cite{mccracken2016fred}. \texttt{Trans} indicates the transformation applied to each variable with $(1)$ implying no transformation, $(2)$ denoting year-on-year growth rates, $(3)$ denoting quarter-on-quarter growth rates, and $(4)$ refers to quarter-on-quarter percentage changes.  
\end{minipage}
\end{center}}
\end{table}

\end{document}